\documentclass[aip,apl,preprint,unsortedaddress]{revtex4-1}
\usepackage{graphicx}
\usepackage{dcolumn}
\usepackage{bm}

\begin{document}

\title{A model for the Pockels effect in distorted liquid crystal blue phases}

\author{F.~Castles}
\affiliation{Department of Materials, University of Oxford, Parks Road, Oxford OX1 3PH, United Kingdom}

\begin{abstract}
Recent experiments have found that a mechanically distorted blue phase can exhibit a primary linear electro-optic (Pockels) effect [F. Castles \textit{et al}. Nature Mater. \textbf{13}, 817 (2014)].  Here it is shown that flexoelectricity can account for the experimental results and a model, which is based on continuum theory but takes account of the sub-unit-cell structure, is proposed.  The model provides a quantitative description of the effect accurate to the nearest order of magnitude and predicts that the Pockels coefficient(s) in an optimally-distorted blue phase may be two orders of magnitude larger than in lithium niobate.
\end{abstract}


\maketitle

The discovery and application of new switching modes in liquid crystal devices has driven the evolution of flat-panel displays from the simple digital wrist watches of the 1970s and 1980s to today's ubiquitous high-definition wide-screen televisions.\cite{handbookv8}  Much recent research has focused on the electro-optic Kerr effect in blue phase (BP) liquid crystals; most notably, Samsung Electronics have used the effect to produce a prototype television.\cite{samsung}  Thus, electro-optic phenomena in liquid crystals in general, and in BPs in particular, are currently of considerable applied interest.  It was recently discovered that a mechanically distorted BP can exhibit a primary linear electro-optic (Pockels) effect in addition to the well known Kerr effect.\cite{castlesnatmat2}  This is of significance in the context of low voltage electro-optic devices because the linear effect necessarily dominates the quadratic Kerr effect at low electric field strengths.  However, the origin of the effect was not determined in Ref.~\onlinecite{castlesnatmat2}.  The purpose of this Letter is to propose a theoretical model for the Pockels effect in mechanically distorted BPs which can account for the previously reported experimental observations.

BPs can exist in materials composed of small elongated organic molecules in a temperature range just below the isotropic liquid phase.\cite{wright}  Of the three BPs that are presently known, the two lowest temperature mesophases---denoted BPI and BPII---exhibit three-dimensional periodicity in the field describing the average orientation of the molecules' long axes, whereas the highest temperature mesophase---denoted BPIII---appears amorphous.\cite{wright}  BPI and BPII, which are the focus of this Letter, belong to the cubic $\bm O(432)$ crystal class,\cite{kitzerow3} and may be referred to collectively as the \textit{cubic BPs}.  The structures, and hence the optical properties, of the cubic BPs are known to be readily modified by an applied electric field via a number of mechanisms, including field-induced phase changes,\cite{armitage,kitzerow} electrostriction \cite{heppke2,kitzerow}, and a direct electro-optic Kerr effect.\cite{porsch,kitzerow}  These effects are conventionally interpreted in terms of dielectric coupling, though they have also been considered as a result of flexoelectric coupling.\cite{kitzerow3,alexander3,alexanderthesis,tiribocchi,outram2}  It is the Kerr effect that has been of particular interest in the context of display devices because it operates on a time scale of $10^{-5}$--$10^{-4}$~s, which is much quicker than the time scale of field-induced phase changes and electrostriction.\cite{gerber,kikuchi,hisakado}  Further, the effective Kerr coefficient in the BPs is remarkably large compared to conventional materials: values $10^2$--$10^4$ times larger than for nitrobenzene have been reported.\cite{hisakado,tian,chen6}

The Pockels effect, despite being of considerable technological importance in other materials,\cite{goldstein} is not usually considered in the context of the BPs.  This is because the symmetries of the undistorted BP structures are such that the Pockels effect is forbidden.\cite{porsch}  Indeed, it was only in a mechanically distorted BP that the Pockels effect was observed;\cite{castlesnatmat2} the distortion reduced the symmetry of the structure to that for which a Pockels effect is permitted.  Apart from providing a more complete description of this new physical phenomenon, a theoretical description the effect will be useful if one seeks to engineer a BP material with an optimized Pockels response for use in devices.

The Pockels effect in a mechanically distorted cubic BP may be readily described within the framework of classical crystal optics by defining a dielectric tensor that is averaged over the BP unit cell.  In this case, the linear electro-optic behavior is characterized by the Pockels tensor (similar to the way in which the quadratic electro-optic behavior is characterized by the Kerr tensor, often approximated as a single Kerr constant\cite{porsch,kitzerow3,gerber,hisakado}).  However, while this macroscopic approach is of great utility, it says nothing about the underlying physical origin of the effect.  The natural approach for a detailed analysis of the BPs is a `mesoscopic' one, which uses continuum theory but takes account of the sub-unit-cell structure.\cite{wright}  The physical origin of the effect is of some interest because it should be entirely different from solid crystals and  its magnitude will not be subject to the same restrictions.  Further, the physical origin must be somewhat exotic even within the context of liquid crystal switching because the experimental results cannot be accounted for by conventional reorientation of the director due to dielectric anisotropy.

There are two main mechanisms by which an external electric field ${\bf E}$ can couple to the liquid crystal director ${\bf n}$: conventional dielectric coupling and flexoelectric coupling.\cite{gennesbook}  The conventional dielectric coupling is described by a term in the free energy density that is $\propto ({\bf E}\cdot{\bf n})^2$, and its effect must therefore be independent of the polarity, or sign, of ${\bf E}$.  This cannot produce the linear electro-optic switching reported in Ref.~\onlinecite{castlesnatmat2}, where the transmitted intensity between crossed polarizers could be different for different polarities of ${\bf E}$.  On the other hand, flexoelectric coupling,\cite{meyer} which is described by a term in the free energy density that is $\propto {\bf E}$, can produce effects that depend on the polarity of the applied field; for example, a linear electro-optic effect due to flexoelectricity is known in the chiral nematic mesophase.\cite{patel}  Therefore one is led to consider flexoelectricity as a mechanism for the Pockels effect in mechanically distorted BPs.

Extensive theoretical investigations concerning the effect of flexoelectric coupling on the structure of the cubic BPs have been reported in Refs.~\onlinecite{alexander3,alexanderthesis,tiribocchi}, and the contribution of flexoelectricity to the Kerr effect has been experimentally and theoretically treated in Ref.~\onlinecite{outram2}.  Kitzerow has theoretically considered the possibility of \textit{polar} flexoelectric electro-optic switching in the cubic BPs in Ref.~\onlinecite{kitzerow}; he argues that, while flexoelectricity leads to local director distortions, the polar optical responses from different parts of the cubic structure should largely cancel.  He concludes that ``the flexoelectric effect in BPs is expected to have little effect on the field-induced birefringence, if any".\cite{kitzerow3}  This is consistent with the known absence of a Pockels effect in the BPs according to classical crystal optics.  However, Kitzerow's analysis naturally concerns BPs that are undistorted in the absence of the electric field.  To develop a model for the Pockels effect in cubic BPs that are mechanically distorted in the ${\bf E}={\bm 0}$ state, I herein build upon and extend Kitzerow's approach.

\begin{figure}
\includegraphics[width=0.43\textwidth]{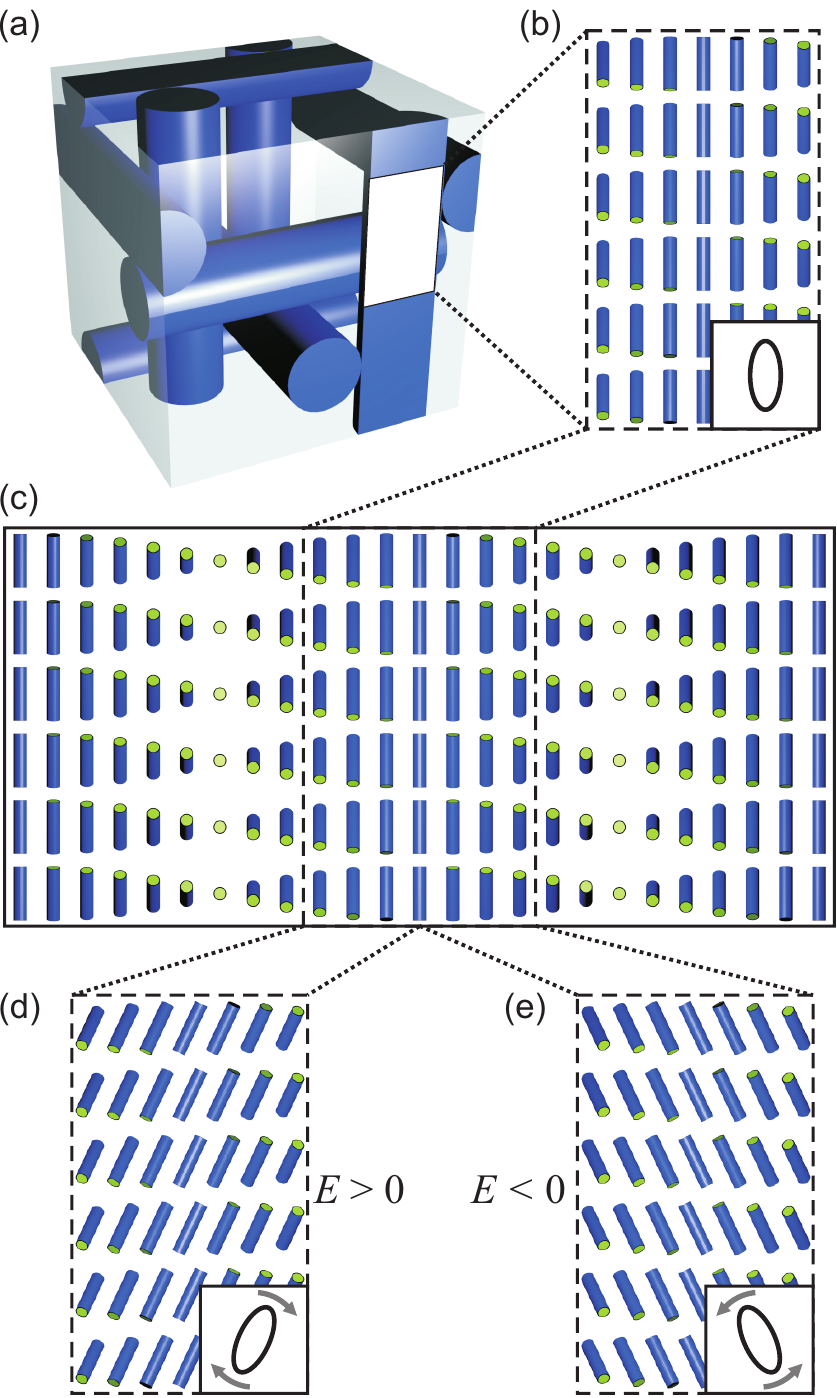}
\caption{\label{fig:singleflexodtc}  (a) Schematic diagram of the DTCs from which BPI may ideally be considered to be composed.\cite{meiboom,wright}  (b) Director pattern in a cross section of a DTC, where the orientation of the cylinders denotes the orientation of ${\bf n}$.  Inset: the effective optical indicatrix associated with the DTC.  (c) Director pattern for a chiral nematic mesophase in a plane which contains the helical axis.  (d)\&(e)  Director distortions in a chiral nematic mesophase under an electric field applied perpendicular to the plane of the diagram.\cite{patel}  The situation is depicted where $P_0>0$, $(e_1-e_3)>0$, and $E>0$ corresponds to a field directed into the page.  Insets: effective optical indicatrices associated with the field-distorted structures.}
\end{figure}

Kitzerow considered\cite{kitzerow3} the effect of the external electric field on the double-twist cylinders (DTCs) from which the cubic BPs may ideally be considered to be composed.\cite{meiboom,wright}  The well known DTC arrangement for BPI is shown in Fig.~\ref{fig:singleflexodtc}(a).  For an undistorted BP subject to flexoelectric coupling, Kitzerow argued that:  (1) An electric field applied along the axis of a DTC will distort the director, but such a distortion will not affect the birefringence.  (2) An electric field perpendicular to a DTC axis will modify its effective birefringence, but this is likely to be canceled by another DTC which is also perpendicular to the field.\cite{kitzerow3}  It remains to consider in more detail the effect of flexoelectricity on the optical properties of a DTC for a component of the electric field that is perpendicular to the DTC axis, and to extend this picture to a mechanically-distorted BP.

The director field in a plane that passes through the axis of a DTC is shown in Fig.~\ref{fig:singleflexodtc}(b).
This pattern is effectively the same as that for a portion of the chiral nematic mesophase in a plane that contains the helical axis, Fig.~\ref{fig:singleflexodtc}(c).  Patel and Meyer\cite{patel} showed that, assuming equality of the splay and bend elastic coefficients ($K_1=K_3\equiv K$), an electric field applied perpendicular to the helical axis of a chiral nematic causes the director to rotate uniformly about the axis defined by the field direction, Fig.~\ref{fig:singleflexodtc}(d)\&(e).  For small $E$, the angle of rotation is given by\cite{patel}
\begin{equation} \label{eq:Ntilt}
\theta^\textnormal{\tiny \textsc{N$^*$}}\approx\frac{P_0}{2\pi}\frac{e}{K}E.
\end{equation}
Here, $e\equiv (e_1-e_3)/2$, where $e_1$ and $e_3$ are the splay and bend flexoelectric coefficients respectively (using the notation of Ref.~\onlinecite{gennesbook}), and $P_0$ is the helicoidal pitch of the chiral nematic.  By analogy, for an electric field component that is perpendicular to the axis of a DTC, one may expect the director in the plane shown in Fig.~\ref{fig:singleflexodtc}(b) to rotate in a somewhat similar fashion.  This general picture is confirmed in the Supplemental Material\cite{suppPRL} using a Landau--de Gennes analysis of BPII attributable to Alexander.\cite{alexanderthesis}  Although the director distortion in the DTC is more complex than the chiral nematic case, the net effect in both cases is to rotate the average orientation of the director about an axis defined by the perpendicular component of the electric field.  The magnitude of the average rotation in the DTC is reduced with respect to the chiral nematic case by a factor of three.\cite{suppPRL}  Thus, for an electric field perpendicular to the axis of a DTC, the average rotation of the director may be written
\begin{equation} \label{eq:BPtilt}
\theta^\textnormal{\tiny \textsc{DTC}}\sim\frac{a}{6\pi}\frac{e}{K}E,
\end{equation}
where the effective `pitch' of the BP is given, to the nearest order of magnitude, by the length of the unit cell $a$.

One may consider the macroscopic optical properties of a given set of parallel DTCs by associating with them an average dielectric tensor $\bm\varepsilon$.  Ignoring spatial dispersion and absorption, $\bm\varepsilon$ is real and symmetric\cite{landaubook} and defines an associated ellipsoid---the optical indicatrix---via the equation $\varepsilon_{ij}^{-1}x_ix_j=1$.  For typical materials, an undistorted DTC will be effectively positively uniaxial with the long axis of the indicatrix aligned along the DTC axis, as shown schematically in the inset to Fig.~\ref{fig:singleflexodtc}(b).  Optically, the result of the net director rotation induced by the field will be to rotate this associated indicatrix, analogous to the chiral nematic case\cite{patel} indicated in Fig.~\ref{fig:singleflexodtc}(d)\&(e).

\begin{figure}
\includegraphics[width=0.45\textwidth]{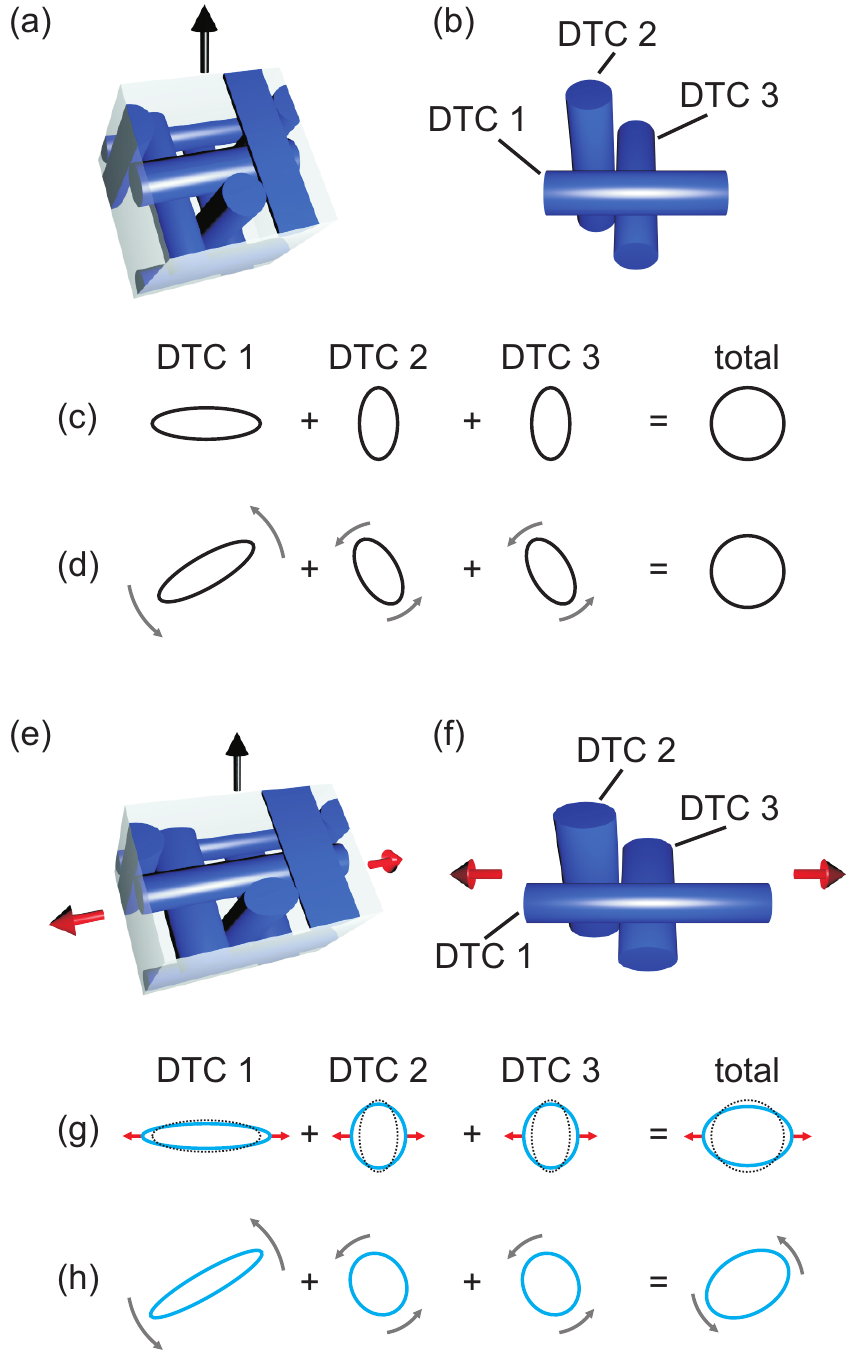}
\caption{\label{fig:flexodtc}  Schematic diagram of flexoelectrically-induced modifications of the optical properties of the double twist cylinders in a cubic BP.  Tiles (a)-(d) concern an undistorted state.  (a) The electric field is applied along a two-fold rotation axis. (b) The principal double twist cylinders, viewed looking back along the electric field direction.  (c) With no electric field, the optical indicatrices associated with each set of double twist cylinders combine to give a nonbirefringent state. (d) With the electric field applied the individual optical indicatrices rotate, but their combined effect remains nonbirefringent, as noted in Ref.~\onlinecite{kitzerow3}.  Tiles (e)-(h) concern a mechanically distorted state.  Red arrows denote the strain. (g) The zero electric field state is now anisotropic. (h) The flexoelectric rotation now causes rotation of the net optic axis.}
\end{figure}

In Fig.~\ref{fig:flexodtc}, a schematic model of the flexoelectric electro-optic effect in BPI is built up by considering the combined effect of the reorientation of the indicatrices associated with each set of parallel DTCs.
  The case of ${\bf E}$ and the viewing direction parallel to a two-fold rotation axis of BPI is considered, corresponding to the experimental configuration investigated in Ref.~\onlinecite{castlesnatmat2}.  As shown in Fig.~\ref{fig:flexodtc}(a)\&(b), in this configuration the electric field is orthogonal to one set of DTCs and at 45$^\circ$ to the other two.  The slices through the associated indicatrices perpendicular to the viewing direction are shown in Fig.~\ref{fig:flexodtc}(c).  For an initially undistorted cubic BP with no field applied, the indicatrices of each DTC combine to form an overall state that is nonbirefringent, as must be the case for a cubic structure in the absence of spatial dispersion.  Upon application of an electric field, Fig.~\ref{fig:flexodtc}(d), the individual indicatrices rotate, but nevertheless again combine to form a nonbirefringent state---as per Kitzerow's account\cite{kitzerow3} and the symmetry arguments of classical crystal optics.\cite{porsch,nyebook}

The equivalent analysis for a mechanically distorted state is shown in Fig.~\ref{fig:flexodtc}(e)--(h).  (In the experiments,\cite{castlesnatmat2} the orientation of the stress with respect to the cubic lattice was typically undetermined; the simplest example, where the stress is along a four-fold rotation axis, is considered in the Figure.)  In the absence of an electric field, the indicatrices now combine to describe an anisotropic state, Fig.~\ref{fig:flexodtc}(g), consistent with the strain-induced breaking of the cubic symmetry and with the experimentally observed birefringence.\cite{castlesnatmat2}  Upon application of an electric field, the net effect is a rotation of the axis of the distortion-induced anisotropy, Fig.~\ref{fig:flexodtc}(h), which is consistent with the experimental observations reported in the Supplementary Information of Ref.~\onlinecite{castlesnatmat2}.

To the nearest order of magnitude, we may approximate that the net rotation of the optic axis in the distorted BP is given by Eq.~(\ref{eq:BPtilt}).  (In fact, the net rotation may be expected to be somewhat lower than this because the field is not oriented optimally perpendicular to every DTC in the BP.  The magnitude of the reduction will probably depend on the specific orientation of the field with respect to the cubic lattice, but it will not be so large as to effect an order of magnitude change and, commensurate with the approximate approach considered herein, this reduction is ignored.)  Inserting into Eq.~(\ref{eq:BPtilt}) the typical approximate values $e\sim 10$~pC/m, $K\sim 10$~pN, $a\approx 300$~nm, and $E\sim 1$~V/$\mu$m, one obtains $\theta\sim 1^\circ$, which is consistent with the experimental observations.

The situation shown in Fig.~\ref{fig:flexodtc} is described within the framework of classical crystal optics as follows.  Uniaxial stress along a four-fold rotation axis causes the symmetry to be reduced $\bm O\rightarrow {\bm D_4}\textrm(422)$.  The Pockels tensor ${\bf r}$, which is defined by $\Delta\varepsilon_{ij}^{-1}=r_{ijk}E_k$, has, for ${\bm D_4}$ symmetry, the nonzero matrix components $r_{41}=-r_{52}$.\cite{glazer}  Setting ${\bf E}$ along the ${\bf x}_1$ axis, and the stress along the ${\bf x}_3$ axis, the equation for the optical indicatrix is
\begin{equation}
\left(x_1, x_2, x_3\right)
\left(
\begin{array}{ccc}
\frac{1}{n_\textnormal{\tiny o}^2}	&	0												&	0 \\
0													&	\frac{1}{n_\textnormal{\tiny o}^2}&	r_{41}E	\\
0													&	r_{41}E									&	\frac{1}{n_\textnormal{\tiny e}^2}	
\end{array}
\right)
\left(
\begin{array}{c}
x_1\\
x_2\\
x_3
\end{array}
\right)
=1
\end{equation}
where $n_\textnormal{\scriptsize o}$ and $n_\textnormal{\scriptsize e}$ are the ordinary and extraordinary refractive indices respectively of the uniaxial ${\bm D_4}$ structure in the absence of an external electric field.  For small $E$, the presence of the electric field term describes a rotation of the indicatrix about the ${\bf x}_1$ axis by an angle
\begin{equation} \label{eq:thetamacroscopic}
\theta=\frac{1}{\frac{1}{n_\textnormal{\tiny o}^2}-\frac{1}{n_\textnormal{\tiny e}^2}}r_{41}E
\approx \frac{n^3}{2\Delta n}r_{41}E,
\end{equation}
where $\Delta n\equiv n_\textnormal{\scriptsize e}- n_\textnormal{\scriptsize o}$ is the birefringence and, to arrive at the expression on the right, it is approximated that $n_\textnormal{\scriptsize o}= n_\textnormal{\scriptsize e}\equiv n$.

Equating the rotations in Eq.~(\ref{eq:BPtilt}) and Eq.~(\ref{eq:thetamacroscopic}) gives
\begin{equation} \label{eq:rBP}
r^\textnormal{\tiny \textsc{BP}}\sim \frac{a\Delta n}{3\pi n^3}\frac{e}{K}.
\end{equation}
This expression links the mesoscopic and macroscopic pictures and provides a metric by which the effectiveness of mechanically distorted BPs as linear electro-optic materials may be assessed.  For the approximate typical values given above (supplemented with $n\approx 1.6$), one finds $r^\textnormal{\tiny \textsc{BP}}\sim \Delta n\times 10^{-8}$~m/V. In the experiments reported in Ref.~\onlinecite{castlesnatmat2} it was determined that $\Delta n\sim 10^{-3}$, which gives $r_\textnormal{\tiny \textsc{BP}}\sim 10^{-11}$~m/V.  This may be compared with LiNbO$_3$---which is reported to be the most widely used linear electro-optic material in industry---for which the nonzero matrix elements of ${\bf r}$ range from 0.7$\times10^{-11}$~m/V to 3$\times10^{-11}$~m/V.\cite{glazer}  If the BP sample were substantially deformed prior to applying the electric field, in which case $\Delta n\sim 0.1$ is reasonable, then an order of magnitude of $r^\textnormal{\tiny \textsc{BP}}\sim 10^{-9}$~m/V appears to be achievable.  Thus, just as the Kerr coefficient(s) in liquid crystal BPs are remarkably large compared to conventional materials, so too may the Pockels coefficient(s) be remarkably large; up to $\sim 10^2$ times larger than for LiNbO$_3$.  Extensive knowledge exists concerning the molecular engineering of the elastic coefficients and flexoelectric coefficients of liquid crystals which, using Eq.~(\ref{eq:rBP}), may now be employed to further optimize the Pockels effect in distorted BPs.

In the above, I have considered the specific case of BPI with the viewing direction and electric-field direction along a two-fold rotation axis, since this was the experimental configuration investigated in Ref.~\onlinecite{castlesnatmat2}.  However, the model may be readily generalized to both cubic BPs and to any stress and electric field directions, yet remain quantitatively accurate to within an order of magnitude or so.  (The corresponding analysis according to classical crystal optics would be modified in such cases; for example, a uniaxial stress along a two-fold rotation axis leads to a biaxial state with ${\bm D_2}$(222) symmetry, for which three elements of the Pockels matrix are nonzero and generally unequal.\cite{glazer})

The model is crude in many respects and assumes a number of quite severe approximations: not least, it ignores completely the regions in between the DTCs.  Further, it ignores spatial dispersion\cite{landaubook} and, as a result, is strictly only valid in the limit that the wavelength of electromagnetic radiation may be considered to be infinite with respect to the lattice dimension; in particular, the approach ignores natural optical activity,\cite{wright,landaubook} the inherent anisotropy of the cubic structure,\cite{landaubook,belyakov3both,demikhov} linear electro-optic effects other than the Pockels effect which exist on account of spatial dispersion,\cite{zheludev1976both,castlespreprintO432} and all aspects relating to selective reflection.  Nevertheless, the model adequately establishes that flexoelectricity can account for the observation of the Pockels effect in mechanically distorted BPs, provides an intuitive picture of the physical origin of the effect in terms of given director distortions, and links the fundamental physical parameters ($K_1$, $K_3$, $e_1$, $e_3$, etc.) with the macroscopic optical properties.

\medskip
The author thanks Stephen Morris, Ben Outram, and Steve Elston for useful discussions.

%\bibliography{C:/biblibrary/references}

\begin{thebibliography}{31}%
\makeatletter
\providecommand \@ifxundefined [1]{%
 \@ifx{#1\undefined}
}%
\providecommand \@ifnum [1]{%
 \ifnum #1\expandafter \@firstoftwo
 \else \expandafter \@secondoftwo
 \fi
}%
\providecommand \@ifx [1]{%
 \ifx #1\expandafter \@firstoftwo
 \else \expandafter \@secondoftwo
 \fi
}%
\providecommand \natexlab [1]{#1}%
\providecommand \enquote  [1]{``#1''}%
\providecommand \bibnamefont  [1]{#1}%
\providecommand \bibfnamefont [1]{#1}%
\providecommand \citenamefont [1]{#1}%
\providecommand \href@noop [0]{\@secondoftwo}%
\providecommand \href [0]{\begingroup \@sanitize@url \@href}%
\providecommand \@href[1]{\@@startlink{#1}\@@href}%
\providecommand \@@href[1]{\endgroup#1\@@endlink}%
\providecommand \@sanitize@url [0]{\catcode `\\12\catcode `\$12\catcode
  `\&12\catcode `\#12\catcode `\^12\catcode `\_12\catcode `\%12\relax}%
\providecommand \@@startlink[1]{}%
\providecommand \@@endlink[0]{}%
\providecommand \url  [0]{\begingroup\@sanitize@url \@url }%
\providecommand \@url [1]{\endgroup\@href {#1}{\urlprefix }}%
\providecommand \urlprefix  [0]{URL }%
\providecommand \Eprint [0]{\href }%
\providecommand \doibase [0]{http://dx.doi.org/}%
\providecommand \selectlanguage [0]{\@gobble}%
\providecommand \bibinfo  [0]{\@secondoftwo}%
\providecommand \bibfield  [0]{\@secondoftwo}%
\providecommand \translation [1]{[#1]}%
\providecommand \BibitemOpen [0]{}%
\providecommand \bibitemStop [0]{}%
\providecommand \bibitemNoStop [0]{.\EOS\space}%
\providecommand \EOS [0]{\spacefactor3000\relax}%
\providecommand \BibitemShut  [1]{\csname bibitem#1\endcsname}%
\let\auto@bib@innerbib\@empty
%</preamble>
\bibitem [{\citenamefont {Goodby}\ \emph {et~al.}(2014)\citenamefont {Goodby},
  \citenamefont {Collings}, \citenamefont {Kato}, \citenamefont {Tschierske},
  \citenamefont {Gleeson},\ and\ \citenamefont {Raynes}}]{handbookv8}%
  \BibitemOpen
  \bibinfo {editor} {\bibfnamefont {J.~W.}\ \bibnamefont {Goodby}}, \bibinfo
  {editor} {\bibfnamefont {P.~J.}\ \bibnamefont {Collings}}, \bibinfo {editor}
  {\bibfnamefont {T.}~\bibnamefont {Kato}}, \bibinfo {editor} {\bibfnamefont
  {C.}~\bibnamefont {Tschierske}}, \bibinfo {editor} {\bibfnamefont {H.~F.}\
  \bibnamefont {Gleeson}}, \ and\ \bibinfo {editor} {\bibfnamefont
  {P.}~\bibnamefont {Raynes}},\ eds.,\ \href@noop {} {\emph {\bibinfo {title}
  {Handbook of Liquid Crystals, Vol. 8: Applications of Liquid Crystals}}}\
  (\bibinfo  {publisher} {Wiley-VCH},\ \bibinfo {address} {Weinheim},\ \bibinfo
  {year} {2014})\BibitemShut {NoStop}%
\bibitem [{sam()}]{samsung}%
  \BibitemOpen
  \href@noop {} {}\bibinfo {note} {Samsung Electronics, {``}15 inch blue phase
  mode {LC} display,{"} seminar and exhibition presented at the Society for
  Information Display 2008 International Symposium}\BibitemShut {NoStop}%
\bibitem [{\citenamefont {Castles}\ \emph {et~al.}(2014)\citenamefont
  {Castles}, \citenamefont {Morris}, \citenamefont {Hung}, \citenamefont
  {Qasim}, \citenamefont {Wright}, \citenamefont {Nosheen}, \citenamefont
  {Choi}, \citenamefont {Outram}, \citenamefont {Elston}, \citenamefont
  {Burgess}, \citenamefont {Hill}, \citenamefont {Wilkinson},\ and\
  \citenamefont {Coles}}]{castlesnatmat2}%
  \BibitemOpen
  \bibfield  {author} {\bibinfo {author} {\bibfnamefont {F.}~\bibnamefont
  {Castles}}, \bibinfo {author} {\bibfnamefont {S.~M.}\ \bibnamefont {Morris}},
  \bibinfo {author} {\bibfnamefont {J.~M.~C.}\ \bibnamefont {Hung}}, \bibinfo
  {author} {\bibfnamefont {M.~M.}\ \bibnamefont {Qasim}}, \bibinfo {author}
  {\bibfnamefont {A.~D.}\ \bibnamefont {Wright}}, \bibinfo {author}
  {\bibfnamefont {S.}~\bibnamefont {Nosheen}}, \bibinfo {author} {\bibfnamefont
  {S.~S.}\ \bibnamefont {Choi}}, \bibinfo {author} {\bibfnamefont {B.~I.}\
  \bibnamefont {Outram}}, \bibinfo {author} {\bibfnamefont {S.~J.}\
  \bibnamefont {Elston}}, \bibinfo {author} {\bibfnamefont {C.}~\bibnamefont
  {Burgess}}, \bibinfo {author} {\bibfnamefont {L.}~\bibnamefont {Hill}},
  \bibinfo {author} {\bibfnamefont {T.~D.}\ \bibnamefont {Wilkinson}}, \ and\
  \bibinfo {author} {\bibfnamefont {H.~J.}\ \bibnamefont {Coles}},\ }\href@noop
  {} {\bibfield  {journal} {\bibinfo  {journal} {Nature Mater.}\ }\textbf
  {\bibinfo {volume} {13}},\ \bibinfo {pages} {817} (\bibinfo {year}
  {2014})}\BibitemShut {NoStop}%
\bibitem [{\citenamefont {Wright}\ and\ \citenamefont {Mermin}(1989)}]{wright}%
  \BibitemOpen
  \bibfield  {author} {\bibinfo {author} {\bibfnamefont {D.~C.}\ \bibnamefont
  {Wright}}\ and\ \bibinfo {author} {\bibfnamefont {N.~D.}\ \bibnamefont
  {Mermin}},\ }\href@noop {} {\bibfield  {journal} {\bibinfo  {journal} {Rev.
  Mod. Phys.}\ }\textbf {\bibinfo {volume} {61}},\ \bibinfo {pages} {385}
  (\bibinfo {year} {1989})}\BibitemShut {NoStop}%
\bibitem [{\citenamefont {Kitzerow}(2010)}]{kitzerow3}%
  \BibitemOpen
  \bibfield  {author} {\bibinfo {author} {\bibfnamefont {H.-S.}\ \bibnamefont
  {Kitzerow}},\ }\href@noop {} {\bibfield  {journal} {\bibinfo  {journal}
  {Ferroelectrics}\ }\textbf {\bibinfo {volume} {395}},\ \bibinfo {pages} {66}
  (\bibinfo {year} {2010})}\BibitemShut {NoStop}%
\bibitem [{\citenamefont {Armitage}\ and\ \citenamefont
  {Cox}(1980)}]{armitage}%
  \BibitemOpen
  \bibfield  {author} {\bibinfo {author} {\bibfnamefont {D.}~\bibnamefont
  {Armitage}}\ and\ \bibinfo {author} {\bibfnamefont {R.~J.}\ \bibnamefont
  {Cox}},\ }\href@noop {} {\bibfield  {journal} {\bibinfo  {journal} {Mol.
  Cryst. Liq. Cryst.}\ }\textbf {\bibinfo {volume} {64}},\ \bibinfo {pages}
  {41} (\bibinfo {year} {1980})}\BibitemShut {NoStop}%
\bibitem [{\citenamefont {Kitzerow}(1991)}]{kitzerow}%
  \BibitemOpen
  \bibfield  {author} {\bibinfo {author} {\bibfnamefont {H.-S.}\ \bibnamefont
  {Kitzerow}},\ }\href@noop {} {\bibfield  {journal} {\bibinfo  {journal} {Mol.
  Cryst. Liq. Cryst.}\ }\textbf {\bibinfo {volume} {202}},\ \bibinfo {pages}
  {51} (\bibinfo {year} {1991})}\BibitemShut {NoStop}%
\bibitem [{\citenamefont {Heppke}, \citenamefont {Krumrey},\ and\ \citenamefont
  {Oestreicher}(1983)}]{heppke2}%
  \BibitemOpen
  \bibfield  {author} {\bibinfo {author} {\bibfnamefont {G.}~\bibnamefont
  {Heppke}}, \bibinfo {author} {\bibfnamefont {M.}~\bibnamefont {Krumrey}}, \
  and\ \bibinfo {author} {\bibfnamefont {F.}~\bibnamefont {Oestreicher}},\
  }\href@noop {} {\bibfield  {journal} {\bibinfo  {journal} {Mol. Cryst. Liq.
  Cryst.}\ }\textbf {\bibinfo {volume} {99}},\ \bibinfo {pages} {99} (\bibinfo
  {year} {1983})}\BibitemShut {NoStop}%
\bibitem [{\citenamefont {Porsch}, \citenamefont {Stegemeyer},\ and\
  \citenamefont {Hiltrop}(1984)}]{porsch}%
  \BibitemOpen
  \bibfield  {author} {\bibinfo {author} {\bibfnamefont {F.}~\bibnamefont
  {Porsch}}, \bibinfo {author} {\bibfnamefont {H.}~\bibnamefont {Stegemeyer}},
  \ and\ \bibinfo {author} {\bibfnamefont {K.}~\bibnamefont {Hiltrop}},\
  }\href@noop {} {\bibfield  {journal} {\bibinfo  {journal} {Z. Naturforsch.}\
  }\textbf {\bibinfo {volume} {39a}},\ \bibinfo {pages} {475} (\bibinfo {year}
  {1984})}\BibitemShut {NoStop}%
\bibitem [{\citenamefont {Alexander}\ and\ \citenamefont
  {Yeomans}(2009)}]{alexander3}%
  \BibitemOpen
  \bibfield  {author} {\bibinfo {author} {\bibfnamefont {G.~P.}\ \bibnamefont
  {Alexander}}\ and\ \bibinfo {author} {\bibfnamefont {J.~M.}\ \bibnamefont
  {Yeomans}},\ }\href@noop {} {\bibfield  {journal} {\bibinfo  {journal} {Liq.
  Cryst.}\ }\textbf {\bibinfo {volume} {36}},\ \bibinfo {pages} {1215}
  (\bibinfo {year} {2009})}\BibitemShut {NoStop}%
\bibitem [{ale()}]{alexanderthesis}%
  \BibitemOpen
  \href@noop {} {}\bibinfo {note} {G. P. Alexander, D.Phil. thesis, University
  of Oxford, 2008.}\BibitemShut {Stop}%
\bibitem [{\citenamefont {Tiribocchi}\ \emph {et~al.}(2013)\citenamefont
  {Tiribocchi}, \citenamefont {Cates}, \citenamefont {Gonnella}, \citenamefont
  {Marenduzzo},\ and\ \citenamefont {Orlandini}}]{tiribocchi}%
  \BibitemOpen
  \bibfield  {author} {\bibinfo {author} {\bibfnamefont {A.}~\bibnamefont
  {Tiribocchi}}, \bibinfo {author} {\bibfnamefont {M.~E.}\ \bibnamefont
  {Cates}}, \bibinfo {author} {\bibfnamefont {G.}~\bibnamefont {Gonnella}},
  \bibinfo {author} {\bibfnamefont {D.}~\bibnamefont {Marenduzzo}}, \ and\
  \bibinfo {author} {\bibfnamefont {E.}~\bibnamefont {Orlandini}},\ }\href@noop
  {} {\bibfield  {journal} {\bibinfo  {journal} {Soft Matter}\ }\textbf
  {\bibinfo {volume} {9}},\ \bibinfo {pages} {4831} (\bibinfo {year}
  {2013})}\BibitemShut {NoStop}%
\bibitem [{\citenamefont {Outram}\ \emph {et~al.}()\citenamefont {Outram},
  \citenamefont {Elston}, \citenamefont {Castles}, \citenamefont {Qasim},
  \citenamefont {Coles}, \citenamefont {Chen},\ and\ \citenamefont
  {Lu}}]{outram2}%
  \BibitemOpen
  \bibfield  {author} {\bibinfo {author} {\bibfnamefont {B.~I.}\ \bibnamefont
  {Outram}}, \bibinfo {author} {\bibfnamefont {S.~J.}\ \bibnamefont {Elston}},
  \bibinfo {author} {\bibfnamefont {F.}~\bibnamefont {Castles}}, \bibinfo
  {author} {\bibfnamefont {M.~M.}\ \bibnamefont {Qasim}}, \bibinfo {author}
  {\bibfnamefont {H.}~\bibnamefont {Coles}}, \bibinfo {author} {\bibfnamefont
  {H.-Y.}\ \bibnamefont {Chen}}, \ and\ \bibinfo {author} {\bibfnamefont
  {S.-F.}\ \bibnamefont {Lu}},\ }\href@noop {} {}\bibinfo {note}
  {{a}rXiv:1406.3125v1}\BibitemShut {NoStop}%
\bibitem [{\citenamefont {Gerber}(1985)}]{gerber}%
  \BibitemOpen
  \bibfield  {author} {\bibinfo {author} {\bibfnamefont {P.~R.}\ \bibnamefont
  {Gerber}},\ }\href@noop {} {\bibfield  {journal} {\bibinfo  {journal} {Mol.
  Cryst. Liq. Cryst.}\ }\textbf {\bibinfo {volume} {116}},\ \bibinfo {pages}
  {197} (\bibinfo {year} {1985})}\BibitemShut {NoStop}%
\bibitem [{\citenamefont {Kikuchi}\ \emph {et~al.}(2002)\citenamefont
  {Kikuchi}, \citenamefont {Yokota}, \citenamefont {Hisakado}, \citenamefont
  {Yang},\ and\ \citenamefont {Kajiyama}}]{kikuchi}%
  \BibitemOpen
  \bibfield  {author} {\bibinfo {author} {\bibfnamefont {H.}~\bibnamefont
  {Kikuchi}}, \bibinfo {author} {\bibfnamefont {M.}~\bibnamefont {Yokota}},
  \bibinfo {author} {\bibfnamefont {Y.}~\bibnamefont {Hisakado}}, \bibinfo
  {author} {\bibfnamefont {H.}~\bibnamefont {Yang}}, \ and\ \bibinfo {author}
  {\bibfnamefont {T.}~\bibnamefont {Kajiyama}},\ }\href@noop {} {\bibfield
  {journal} {\bibinfo  {journal} {Nature Mater.}\ }\textbf {\bibinfo {volume}
  {1}},\ \bibinfo {pages} {64} (\bibinfo {year} {2002})}\BibitemShut {NoStop}%
\bibitem [{\citenamefont {Hisakado}\ \emph {et~al.}(2005)\citenamefont
  {Hisakado}, \citenamefont {Kikuchi}, \citenamefont {Nagamura},\ and\
  \citenamefont {Kajiyama}}]{hisakado}%
  \BibitemOpen
  \bibfield  {author} {\bibinfo {author} {\bibfnamefont {Y.}~\bibnamefont
  {Hisakado}}, \bibinfo {author} {\bibfnamefont {H.}~\bibnamefont {Kikuchi}},
  \bibinfo {author} {\bibfnamefont {T.}~\bibnamefont {Nagamura}}, \ and\
  \bibinfo {author} {\bibfnamefont {T.}~\bibnamefont {Kajiyama}},\ }\href@noop
  {} {\bibfield  {journal} {\bibinfo  {journal} {Adv. Mater.}\ }\textbf
  {\bibinfo {volume} {17}},\ \bibinfo {pages} {96} (\bibinfo {year}
  {2005})}\BibitemShut {NoStop}%
\bibitem [{\citenamefont {Tian}\ \emph {et~al.}(2013)\citenamefont {Tian},
  \citenamefont {Goodby}, \citenamefont {G{\"o}rtz},\ and\ \citenamefont
  {Gleeson}}]{tian}%
  \BibitemOpen
  \bibfield  {author} {\bibinfo {author} {\bibfnamefont {L.}~\bibnamefont
  {Tian}}, \bibinfo {author} {\bibfnamefont {J.~W.}\ \bibnamefont {Goodby}},
  \bibinfo {author} {\bibfnamefont {V.}~\bibnamefont {G{\"o}rtz}}, \ and\
  \bibinfo {author} {\bibfnamefont {H.~F.}\ \bibnamefont {Gleeson}},\
  }\href@noop {} {\bibfield  {journal} {\bibinfo  {journal} {Liq. Cryst.}\
  }\textbf {\bibinfo {volume} {40}},\ \bibinfo {pages} {1446} (\bibinfo {year}
  {2013})}\BibitemShut {NoStop}%
\bibitem [{\citenamefont {Chen}\ \emph {et~al.}(2013)\citenamefont {Chen},
  \citenamefont {Xu}, \citenamefont {Wu}, \citenamefont {Yamamoto},\ and\
  \citenamefont {Haseba}}]{chen6}%
  \BibitemOpen
  \bibfield  {author} {\bibinfo {author} {\bibfnamefont {Y.}~\bibnamefont
  {Chen}}, \bibinfo {author} {\bibfnamefont {D.}~\bibnamefont {Xu}}, \bibinfo
  {author} {\bibfnamefont {S.-T.}\ \bibnamefont {Wu}}, \bibinfo {author}
  {\bibfnamefont {S.}~\bibnamefont {Yamamoto}}, \ and\ \bibinfo {author}
  {\bibfnamefont {Y.}~\bibnamefont {Haseba}},\ }\href@noop {} {\bibfield
  {journal} {\bibinfo  {journal} {Appl. Phys. Lett.}\ }\textbf {\bibinfo
  {volume} {102}},\ \bibinfo {pages} {141116} (\bibinfo {year}
  {2013})}\BibitemShut {NoStop}%
\bibitem [{\citenamefont {Goldstein}(1968)}]{goldstein}%
  \BibitemOpen
  \bibfield  {author} {\bibinfo {author} {\bibfnamefont {R.}~\bibnamefont
  {Goldstein}},\ }\href@noop {} {\enquote {\bibinfo {title} {Pockels cell
  primer},}\ }\bibinfo {howpublished} {Laser Focus, p. 21} (\bibinfo {year}
  {February 1968})\BibitemShut {NoStop}%
\bibitem [{\citenamefont {de~Gennes}\ and\ \citenamefont
  {Prost}(1993)}]{gennesbook}%
  \BibitemOpen
  \bibfield  {author} {\bibinfo {author} {\bibfnamefont {P.-G.}\ \bibnamefont
  {de~Gennes}}\ and\ \bibinfo {author} {\bibfnamefont {J.}~\bibnamefont
  {Prost}},\ }\href@noop {} {\emph {\bibinfo {title} {The Physics of Liquid
  Crystals}}},\ \bibinfo {edition} {2nd}\ ed.\ (\bibinfo  {publisher} {Oxford
  University Press, Oxford},\ \bibinfo {year} {1993})\BibitemShut {NoStop}%
\bibitem [{\citenamefont {Meyer}(1969)}]{meyer}%
  \BibitemOpen
  \bibfield  {author} {\bibinfo {author} {\bibfnamefont {R.~B.}\ \bibnamefont
  {Meyer}},\ }\href@noop {} {\bibfield  {journal} {\bibinfo  {journal} {Phys.
  Rev. Lett.}\ }\textbf {\bibinfo {volume} {22}},\ \bibinfo {pages} {918}
  (\bibinfo {year} {1969})}\BibitemShut {NoStop}%
\bibitem [{\citenamefont {Patel}\ and\ \citenamefont {Meyer}(1987)}]{patel}%
  \BibitemOpen
  \bibfield  {author} {\bibinfo {author} {\bibfnamefont {J.~S.}\ \bibnamefont
  {Patel}}\ and\ \bibinfo {author} {\bibfnamefont {R.~B.}\ \bibnamefont
  {Meyer}},\ }\href@noop {} {\bibfield  {journal} {\bibinfo  {journal} {Phys.
  Rev. Lett.}\ }\textbf {\bibinfo {volume} {58}},\ \bibinfo {pages} {1538}
  (\bibinfo {year} {1987})}\BibitemShut {NoStop}%
\bibitem [{\citenamefont {Meiboom}\ \emph {et~al.}(1981)\citenamefont
  {Meiboom}, \citenamefont {Sethna}, \citenamefont {Anderson},\ and\
  \citenamefont {Brinkman}}]{meiboom}%
  \BibitemOpen
  \bibfield  {author} {\bibinfo {author} {\bibfnamefont {S.}~\bibnamefont
  {Meiboom}}, \bibinfo {author} {\bibfnamefont {J.~P.}\ \bibnamefont {Sethna}},
  \bibinfo {author} {\bibfnamefont {P.~W.}\ \bibnamefont {Anderson}}, \ and\
  \bibinfo {author} {\bibfnamefont {W.~F.}\ \bibnamefont {Brinkman}},\
  }\href@noop {} {\bibfield  {journal} {\bibinfo  {journal} {Phys. Rev. Lett.}\
  }\textbf {\bibinfo {volume} {46}},\ \bibinfo {pages} {1216} (\bibinfo {year}
  {1981})}\BibitemShut {NoStop}%
\bibitem [{sup()}]{suppPRL}%
  \BibitemOpen
  \href@noop {} {}\bibinfo {note} {See Supplemental Material at [URL will be
  inserted by publisher]}\BibitemShut {NoStop}%
\bibitem [{\citenamefont {Landau}, \citenamefont {Lifshitz},\ and\
  \citenamefont {Pitaevskii}(1984)}]{landaubook}%
  \BibitemOpen
  \bibfield  {author} {\bibinfo {author} {\bibfnamefont {L.~D.}\ \bibnamefont
  {Landau}}, \bibinfo {author} {\bibfnamefont {E.~M.}\ \bibnamefont
  {Lifshitz}}, \ and\ \bibinfo {author} {\bibfnamefont {L.~P.}\ \bibnamefont
  {Pitaevskii}},\ }\href@noop {} {\emph {\bibinfo {title} {Electrodynamics of
  Continuous Media (Course of Theoretical Physics, Vol. 8)}}},\ \bibinfo
  {edition} {2nd}\ ed.\ (\bibinfo  {publisher} {Butterworth-Heinemann,
  Oxford},\ \bibinfo {year} {1984})\BibitemShut {NoStop}%
\bibitem [{\citenamefont {Nye}(1985)}]{nyebook}%
  \BibitemOpen
  \bibfield  {author} {\bibinfo {author} {\bibfnamefont {J.~F.}\ \bibnamefont
  {Nye}},\ }\href@noop {} {\emph {\bibinfo {title} {Physical Properties of
  Crystals}}}\ (\bibinfo  {publisher} {Oxford University Press, Oxford},\
  \bibinfo {year} {1985})\BibitemShut {NoStop}%
\bibitem [{\citenamefont {Glazer}\ and\ \citenamefont {Cox}(2003)}]{glazer}%
  \BibitemOpen
  \bibfield  {author} {\bibinfo {author} {\bibfnamefont {A.~M.}\ \bibnamefont
  {Glazer}}\ and\ \bibinfo {author} {\bibfnamefont {K.~G.}\ \bibnamefont
  {Cox}},\ }in\ \href@noop {} {\emph {\bibinfo {booktitle} {International
  Tables for Crystallography, Vol. D: Physical Properties of Crystals}}},\
  \bibinfo {editor} {edited by\ \bibinfo {editor} {\bibfnamefont
  {A.}~\bibnamefont {Authier}}}\ (\bibinfo  {publisher} {Kluwer Academic
  Publishers, Dordrecht},\ \bibinfo {year} {2003})\ p.\ \bibinfo {pages}
  {150}\BibitemShut {NoStop}%
\bibitem [{bel()}]{belyakov3both}%
  \BibitemOpen
  \href@noop {} {}\bibinfo {note} {V. A. Belyakov, V. E. Dmitrenko, and S. M.
  Osadchi\u{i}, Zh. Eksp. Teor. Fiz. \textbf{83}, 585 (1982) [Sov. Phys. JETP
  \textbf{56}, 322, (1982)]}\BibitemShut {NoStop}%
\bibitem [{\citenamefont {Demikhov}\ and\ \citenamefont
  {Stegemeyer}(1993)}]{demikhov}%
  \BibitemOpen
  \bibfield  {author} {\bibinfo {author} {\bibfnamefont {E.}~\bibnamefont
  {Demikhov}}\ and\ \bibinfo {author} {\bibfnamefont {H.}~\bibnamefont
  {Stegemeyer}},\ }\href@noop {} {\bibfield  {journal} {\bibinfo  {journal}
  {Liq. Cryst.}\ }\textbf {\bibinfo {volume} {14}},\ \bibinfo {pages} {1801}
  (\bibinfo {year} {1993})}\BibitemShut {NoStop}%
\bibitem [{\citenamefont {Zheludev}()}]{zheludev1976both}%
  \BibitemOpen
  \bibfield  {author} {\bibinfo {author} {\bibfnamefont {I.~S.}\ \bibnamefont
  {Zheludev}},\ }\href@noop {} {\bibfield  {journal} {\bibinfo  {journal} {Usp.
  Fiz. Nauk.}\ }\textbf {\bibinfo {volume} {120}},\ \bibinfo {pages} {702}},\
  \bibinfo {note} {(1976) [Sov. Phys. Usp. \textbf{19}, 1029,
  (1976)]}\BibitemShut {NoStop}%
\bibitem [{\citenamefont {Castles}()}]{castlespreprintO432}%
  \BibitemOpen
  \bibfield  {author} {\bibinfo {author} {\bibfnamefont {F.}~\bibnamefont
  {Castles}},\ }\href@noop {} {}\bibinfo {note}
  {{a}rXiv:1503.04103v2}\BibitemShut {NoStop}%
\end{thebibliography}
%
\end{document}